\documentclass[twocolumn,showpacs,prb]{revtex4}
\usepackage{graphicx}%
\usepackage{dcolumn}
\usepackage{amsmath}

\begin{document}

\title{Unusual behaviors in the transport properties of REFe$_{4}$P$_{12}$ (RE: La, Ce, Pr, and Nd)}

\author{H. Sato, Y. Abe, H. Okada, T. D. Matsuda, K. Abe, H. Sugawara and Y. Aoki}

\affiliation{Department of Physics, Tokyo Metropolitan University, Tokyo 192-0397, Japan}

\date{\today}

\begin{abstract} 
We have investigated the resistivity ($\rho$), thermoelectric power (TEP) and Hall coefficient ($R_{H}$) on high quality single crystals of REFe$_{4}$P$_{12}$.  TEP in CeFe$_{4}$P$_{12}$ is extremely large ($\sim$ 0.5mV/K at 290K) with a peak of $\sim$ 0.75mV/K at around 65K.  The Hall mobility also shows a peak at $\sim$ 65K, suggesting carriers with heavy masses developed at lower temperatures related with the f-hybridized band.  Both Pr- and Nd- systems exhibit an apparent increase of $\rho$ with decreasing temperature far above their magnetic transition temperatures.  In the same temperature ranges, TEP exhibits unusually large absolute values of -50$\mu$V/K for PrFe$_{4}$P$_{12}$ and -15$\mu$V/K for NdFe$_{4}$P$_{12}$, respectively.  For PrFe$_{4}$P$_{12}$, such anomalous transport properties suggest an unusual ground state, possibly related with the Quadrupolar Kondo effect.
\end{abstract}


\maketitle

\section{INTRODUCTION} 
Recently, the rare earth compounds RETr$_{4}$Pn$_{12}$ (Tr: Fe, Co, Ru, Os and Pn: P, As, Sb) with the filled skutterudite structure$^{1-3}$  have attracted much attention, partly because of a potential practical application as thermoelectric materials.$^{4, 5}$  Also from the viewpoint of basic physics, a number of novel phenomena, such as a metal-insulator transition and intermediate valence, have been reported.$^{6-8}$   Torikachvili et al.$^{2}$ have reported the first systematic study on REFe$_{4}$P$_{12}$ where they found variety of interesting properties depending on RE-atoms; i.e., superconducting LaFe$_{4}$P$_{12}$ below 4.2K, insulating CeFe$_{4}$P$_{12}$, antiferromagnetic PrFe$_{4}$P$_{12}$ below 6.4K, and ferromagnetic NdFe$_{4}$P$_{12}$ below 1.9K.  An uranium analog UFe$_{4}$P$_{12}$ was also reported to be insulating.$^{9}$  For a better understanding of these novel phenomena, high quality single crystals are essential.  We have recently succeeded in growing high quality REFe$_{4}$P$_{12}$ single crystals that realized the first de Haas-van Alphen (dHvA) experiment.$^{10}$  In this paper, we report the first experimental studies (except for our preliminary report in the SCES'98 proceeding$^{11}$) on the thermoelectric power and Hall effect of REFe$_{4}$P$_{12}$ single crystals along with the anomalous behaviors in the electrical resistivity.  
\section{EXPERIMENTAL}  
Single crystals were grown by the tin-flux method,$^{1}$ using 99.99\% Fe, La and Pr, 99.9\% Nd, 99.9999\% P and 99.999\% Sn.  The samples for the transport measurements are basically the same as were used for the dHvA measurements,$^{10}$ and both measurements were made on exactly the same samples for some compounds.  Their Dingle temperatures were estimated to be 0.3$\sim$0.6K, which insures the samples of high quality.  The high quality is also reflected in the large residual resistance ratio (RRR) of 1,300 and 1,000 for La- and Nd-samples, respectively.  The small RRR of $\sim$13 for PrFe$_{4}$P$_{12}$ at zero field does not indicate a low sample quality, since the Fermi surface (FS) changes across $T_{N}$.  The intrinsic RRR of PrFe$_{4}$P$_{12}$ was estimated to be $\sim$1,000 from the resistivity at 0.3K in high fields above the metamagnetic transition.  We found a meaningful correlation between the dHvA amplitude and the RRR; i.e., the dHvA oscillations could be detected only in large RRR samples. 
The electrical resistivity and Hall resistivity were measured by the ordinary dc four-probe method.  Temperature was monitored by calibrated two $RuO_{2}$ and a platinum resistance thermometers depending on the temperature ranges between 0.5K and 295K.  The calibration was made against a thin-film resistance temperature sensor (Cernox resistor, Lake Shore) calibrated against the International Temperature scale of 1990 (ITS-90).  In a supplemental cryostat used only above 1.5K, a Au-0.07\%Fe vs. Chromel thermocouple was used as a temperature monitor.  The thermoelectric power was measured by the differential method using Au-0.07\%Fe vs. Chromel thermocouples.  The thermocouples were calibrated after every cooling down at the boiling point of $^{4}$He in ambient pressure.  The accuracy of temperature measurement is less than 0.1K at low temperatures, while it is a few K near room temperature (RT).  The voltage measurements were made by Keithley 182 nanovoltmeters.  The magnetic measurements were made by a Quantum Design SQUID magnetometer up to 5.5 T.

\section{RESULTS AND DISCUSSION} 
\subsection{Electrical resistivity}
The temperature dependence of electrical resistivity $\rho(T)$ for LaFe$_{4}$P$_{12}$ (not shown) almost agrees with the result in ref.2 except the orders of magnitude larger RRR in the present work.  The superconducting transition temperature ($T_{S}$) estimated from the midpoint of the resistive transition is 4.6K with a transition width of $\Delta$T$\sim$0.6K.  The higher $T_{S}$ in the $\rho(T)$ measurement is not attributable to the reported enhancement of $T_{S}$ with pressure in LaFe$_{4}$P$_{12}$,$^{12}$ since $T_{S}$ estimated from a SQUID measurement under almost the same sample condition is 4.17K in agreement with the previous report.$^{2}$  The transition in $\rho(T)$ may not be a bulk property, but is plausibly the surface superconductivity as is occasionally observed in rare earth superconductors.  In fact, $\rho(T)$ for LaFe$_{4}$P$_{12}$ in ref.2 also shows a tendency to decrease at around 7K.  Below 25K, $\rho$ is well described by $\rho(T)$ = $\rho_{0}$+A$_{0}T^{2}$, with A$_{0}$=(1.0$\pm$0.1)$\times10^{-3}$$\mu\Omega$cmK$^{-2}$.   The large A$_{0}$  compared to the ordinary metals reflects a large contribution from the 3d-band of Fe to the electronic density of states at the Fermi level, as was already indicated in the large specific heat coefficient.$^{13}$

Figure~1 shows the temperature dependence of resistivity for PrFe$_{4}$P$_{12}$.
The most remarkable difference in $\rho(T)$ for PrFe$_{4}$P$_{12}$ between the present work and ref.2 is the sign of d$\rho$/d$T$ above the Neel temperature ($T_{N}$).   In Fig.1, $\rho(T)$ increases almost logarithmically with decreasing temperature between 100K and 30K, in contrast with the monotonous decrease from RT down to $T_{N}$ in ref.2.  The absolute value of $\rho(T)$ is slightly sample dependent in the present work due to irregular shapes of the samples.  However, $\rho(T)$s normalized at the peak for four samples from different batches lie almost on a single curve as shown in the inset of Fig. 1.  The negative d$\rho$/d$T$ is reminiscent of the Kondo effect that has been frequently observed in Ce and U compounds.  For Pr-compounds, PrSn$_{3}^{\ 14}$ and PrInAg$_{2}$$^{15}$ are the quite rare examples exhibiting the Kondo behaviors.  A notable feature of PrFe$_{4}$P$_{12}$ compared to these two compounds is a large magnitude of the logarithmic increase in $\rho(T)$.  A large $C_{e}/T$ value ($>$1J/mol $\cdot$ K$^{2}$) reported at low temperatures in the high field paramagnetic state,$^{16}$ indicating the existence of mass-enhanced electrons, is another sign of the Kondo effect in PrFe$_{4}$P$_{12}$.  If the logarithmic increase in $\rho(T)$ is ascribed to the Kondo scattering, the Kondo temperature $T_{K}$ should be $\sim$100K.  The $T_{K}$ value is in conflict with the Weiss temperature ($\theta_{P}$) of a few K that provides a measure of $T_{K}$ for heavy Fermion compounds, though Kondo scattering associated with higher crystal field levels could not be ruled out as an origin of such an enhancement of $T_{K}$ in $\rho(T)$.$^{17}$  The origin of the logarithmic increase in $\rho$ is an attractive subject for further investigations.  

With decreasing temperature, $\rho$ shows a sharp upturn at 6.7K followed by a peak around 5.4K.  The general features agree with the result in ref.2, although the ratio RR=$\rho(7.5K)$/$\rho(0.5K)$=12$\sim$20 in the present experiment shows a sharp contrast with RR$<$1 in ref.2.  A notable feature in Fig.1 is an abrupt change in slope of $\rho(T)$ curves at $\sim$2K, which was already found on one of the samples in ref. 2.  A sudden increase in $\rho(T)$ above 2K suggests some gap structure in the scattering probability.  Tentatively assuming a resistivity component with the form $\rho_{G}$=A$_{1}T^{2}$exp(-$\Delta/k_{B}T$), $^{18}$ $\rho(T)$ below 3.2K is well reproduced by using $\Delta/k_{B}$=6.8K as given in Fig.~2.  
This fact suggests that the scattering probability of conduction electrons reflects some gap structure, probably in the magnon dispersion relation, although the agreement of $\Delta/k_{B}$=6.8K with $T_{N}$ may be accidental.  It should be noted that a bend was observed near 2K also in the temperature dependence of specific heat coefficient.$^{16}$

$\rho(T)$ for NdFe$_{4}$P$_{12}$ is in agreement with that in ref. 2 except the low residual resistivity ($\le 1/20$) in the present experiment.  Figure~3 shows the low temperature part of $\rho(T)$ for NdFe$_{4}$P$_{12}$ where d$\rho$/d$T$ is negative between $\sim$4K and 30K. 
The negative d$\rho$/d$T$ developing below $\sim$10$\times$$T_{C}$ is quite unusual for ferromagnetic materials. The minimum should be intrinsic, since it has been reproducibly observed in samples with large RRR values.   Below a sharp drop at $T_{C}\sim$2K, the exponent $\it N$ in $\rho(T) =\rho_{0}$ +A$_{0}T^{N}$ is close to 4 as shown in the inset.  The exponent, twice as large as  $\it N$=2 expected for the simple magnon scattering, is well correlated with the $T^{3}$ dependence of specific heat reported in ref. 2 (in contrast with the $T^{3/2}$ dependence expected for simple ferromagnets).  Such large exponents for both the resistivity and specific heat could be consistently understood,$^{19}$  if the magnon dispersion relation is $\hbar$$\omega$$\propto q $ rather than $\hbar\omega \propto q^{2}$.

Figure~4 shows $\rho(T)$ in CeFe$_{4}$P$_{12}$ for four different samples.
The resistivity change between RT and the lowest temperature in the present experiment is about five orders of magnitude smaller than that in ref. 2.  It should be noted again that dHvA signals with large effective masses have been detected in La-, Pr- and Nd-Fe$_{4}$P$_{12}$$^{10,20}$ grown by the same procedure as described above, insuring the quality of the bulk part of the samples.  We thus infer that the effect of impurity doping in the bulk part of samples might be less in the present work.  Taking into account these facts, some shortening effect by small size grains of Sn inevitably embedded in the sample is the most probable origin for such difference in $\rho(T)$ between the two works.  We have tried to reduce the embedded Sn grains by following two ways, (1) repeatedly slicing and etching a sample in acid to dissolve Sn, and (2) selecting samples from different crystals from different batches.  From the Meissner diamagnetic contribution to the low field magnetic susceptibility, we have found the amount of Sn varying from $\sim$1 to less than 0.01-vol.\%.  However, despite the large variation in the Sn content, only minor change in $\Delta \rho/\rho_{280K}$=($\rho_{1.5K}-\rho_{280K})/\rho_{280K}$ was observed.  Note that a large sample-to-sample variation in the conducting behavior in CeFe$_{4}$P$_{12}$ was already reported in refs. 3 and 18, although the $\rho(T)$ curve was shown only for a single sample.  It should be also noted that the resistivity at RT in the present work is slightly larger than that estimated in ref.21. 

The temperature dependence of $\rho$ in Fig. 4 indicates the existence of three characteristic temperature ranges; i.e., T$>$230K, 230K$>$T$>$50K and T$<$50K.  The energy gap 1,250K estimated from the $\rho(T)$ data above 250K is almost sample-independent. The value is close to the reported value of 1,500K in ref. 3 which was estimated in the temperature range 85K$<$T$<$140K.  These values are about 1/3 of the gap (=0.34eV) estimated by the band structure calculation within the local-density approximation.$^{22}$  Taking into account the fact that a recent optical experiment also gives a smaller gap of 0.15eV,$^{21}$ the disagreement may be ascribed to the local density approximation that tends to overestimate the band gaps.$^{22}$ $\rho(T)$ depends only slightly on temperature from 230K to $\sim$50K below which it shows a sharp upturn.  The upturn below 50K, which is hardly explainable if the shortening effect of metallic Sn grains dominates the conduction, is probably an intrinsic behavior of $\rho(T)$.  At this stage, we cannot give any conclusive remark on the origin of the difference in $\rho(T)$ behaviors in the two works.  However, it might be worthwhile to report the first Hall effect and thermoelectric power experiments on CeFe$_{4}$P$_{12}$, since the main characteristics are reproducible.

\subsection{Hall effect}
Figure~5 shows the temperature dependence of Hall coefficient $R_{H}$ for (a) La-, Pr- Nd- Fe$_{4}$P$_{12}$ and (b) CeFe$_{4}$P$_{12}$.
At RT, $R_{H}$ is $\sim$1.5$\times 10^{-9}$m$^{3}$/C for both La- and Nd-Fe$_{4}$P$_{12}$, and $\sim$1.3$\times 10^{-9}$m$^{3}$/C for PrFe$_{4}$P$_{12}$. The Hall coefficient in magnetic system is usually expressed as a combination of the normal Hall coefficient $R_{H0}$ resulting from the Lorentz motion of conduction electrons and the extraordinary one $R_{ex}$ caused by magnetic scattering of conduction electrons; $R_{H}=R_{H0} + R_{ex}$.$^{23-25}$  The positive $R_{H}$  at RT, where $R_{H0}$ dominates, is consistent with the band structure calculation for LaFe$_{4}$P$_{12}$ predicting two hole-like Fermi sheets.$^{26}$  The weak temperature dependence of $R_{H}$ for LaFe$_{4}$P$_{12}$, including a peak at ~25K, can be understood as due to the temperature dependence of the anisotropy in the relaxation time $\tau({\it \bf k})$.$^{27}$  As shown in the inset of Fig. 5(a), the change in $R_{H}$ below $T_{N}$ for PrFe$_{4}$P$_{12}$ is more than two orders of magnitude larger than that above $T_{N}$, which also indicates a FS reconstruction associated with the antiferromagnetic superzone gap formation.  $R_{H}(T)$ becomes almost temperature independent below 2K, which also suggests some gap structure described above in correlation with the anomalies in $\rho(T)$ and the specific heat.$^{16}$

At temperatures higher than Kondo temperature,$^{24,25}$ $R_{ex}$ in the dense Kondo compounds has been well described by the skew scattering mechanism as $R_{ex}=\gamma \chi_{red}\times\rho_{m}$, using the reduced magnetic susceptibility $\chi_{red}$, the magnetic part of resistivity $\rho_{m}$ and a parameter $\gamma$ depending on the phase shift associated with the Kondo scattering.  To test the model, $R_{H}$ is compared with M$\times\rho$ calculated from the magnetization at 0.1T and zero field resistivity for PrFe$_{4}$P$_{12}$  as a function of temperature in Fig.~6 (a).  
$R_{H}(T)$ is qualitatively reproduced by the $M\times\rho$ curves above $T_{N}$.  In contrast, the experimental curve for NdFe$_{4}$P$_{12}$ is hardly reproduced over the wide temperature range, even if the temperature dependence of $R_{H0}$ is taken into account.  This fact suggests that the skew scattering contribution is small at high temperatures.  Thus, the comparison between $R_{H}$ and $M\times\rho$ is shown in Fig. 6 (b) only near $T_{C}$, where the extraordinary Hall resistivity $R_{S}M$ is expected to become larger.$^{22}$  The positive peak is well reproduced as a combined effect of growing M and decreasing $\rho$ below $T_{C}$, though the peak for $R_{H}$ is broader and shifted to higher temperatures.  The disagreement could be partly ascribed to the higher applied field of 1T in the $R_{H}$ measurement.

For CeFe$_{4}$P$_{12}$, despite the clear difference in the $\rho(T)$ curves, $R_{H}(T)$s for the two different samples ($\#$1 and $\#$2) are basically same including the absolute values, suggesting the overall feature of $R_{H}$ to be a bulk property.  Corresponding to the hump in $\rho(T)$, a change in the slope of $R_{H}(T)$ exists around 220K.  A notable feature in Fig. 5(b) is a sharp drop of $R_{H}$ below a peak at $\sim$25K in correlation with the low temperature upturn in $\rho(T)$.  Below RT, the sign is always positive, suggesting the higher mobility carriers in the top valence band.  Of course, the existence of acceptor type impurities could not be ruled out at this stage.  Assuming the single carrier model, the concentration of carriers is estimated to be 1$\times10^{-3}$holes/f.u. at RT, 1$\times10^{-4}$holes/f.u. at 25K (peak in $R_{H}$) and 4$\times10^{-4}$holes/f.u. at 2K. 

The combination of decreasing $R_{H}$ and increasing $\rho(T)$ with decreasing T leads to a decrease of Hall mobility below 25K based on the single carrier model as shown in Fig.~7.
The Hall mobility at RT is about an order of magnitude smaller than that in Si.$^{28}$  The decrease in Hall mobility above $\sim$150K with increasing T follows $T^{-3/2}$ dependence, suggesting the dominant acoustic phonon scattering as expected for ordinary semiconductors.$^{28}$   Note the apparent peak in the Hall mobility observed around 65K.  

\subsection{Thermoelectric power}
Figure~8 (a) shows the temperature dependence of TEP for La-, Pr-, and Nd-compounds.  
TEP for LaFe$_{4}$P$_{12}$ decreases with decreasing T from $\approx$9$\mu$V/K at 290K, and changes its sign below 18K after showing a minimum near 200K and a maximum near 100K.  The relatively large absolute value at RT compared to simple metals is ascribable to the large contribution from 3d band of Fe at Fermi surface.$^{26,29)}$.  A delicate combination of the diffusion and the phonon drag contributions might lead to such a complex T-dependence.

TEP for NdFe$_{4}$P$_{12}$ is almost indistinguishable with that of LaFe$_{4}$P$_{12}$ above 220K, suggesting a localized nature of 4f-electrons in this system.  At lower temperatures, TEP deviates from that for LaFe$_{4}$P$_{12}$ downward and changes its sign at around 85K. At the lowest temperature, it approaches to zero after showing a minimum of $\sim$15$\mu$V/K which is quite large as a localized f-electron material.  The coincidence of the minimum temperature ($\sim$30K) with that for the $\rho(T)$-minimum suggests that these minima originate from the same conduction-electron scattering with a characteristic energy scale of $\sim$30K.

TEP for PrFe$_{4}$P$_{12}$ shows largely different behaviors.  Even at 290K, the magnitude is about twice as large as that in LaFe$_{4}$P$_{12}$, and decreases with decreasing T.  Below a sign change near 70K, it drastically decreases with T, showing a bend near $T_{N}$.  After exhibiting a large minimum near 3K, it approaches to 0$\mu$V/K at the lowest temperature.  The sharp drop below $T_{N}$ could be ascribed to the FS reconstruction.  However, the large magnitude ($\sim$50$\mu$V/K) above $T_{N}$ is quite unusual.  

The diffusion thermoelectric power is represented by an energy derivative of conductivity $\sigma(\varepsilon)$ as,$^{30}$

\[
 S = -(\pi^{2}k_{B}^{2}T/3e\sigma)\{d\sigma(\varepsilon)/d\varepsilon \}_{\varepsilon_{F} }  
\].

The origins of energy dependence of $\sigma(\varepsilon)$  could be further classified into two factors; carrier numbers and scattering probability.  In the high mobility semiconductors where the former factor is huge, a large value of TEP could be expected at low temperatures.  It is not applicable to PrFe$_{4}$P$_{12}$, since we already know the existence of rather large FS in PrFe$_{4}$P$_{12}$ from the dHvA experiments.$^{20}$  Even in the transition metals and alloys with a large electronic density of states near Fermi energy, a large diffusion TEP has been reported only at high temperatures.$^{31}$  In fact, this is the case for TEP in LaFe$_{4}$P$_{12}$.  Such a large magnitude of TEP in metals with a large FS has been reported only for the Kondo systems where a large energy dependent scattering process is essential.$^{31,32}$  If we assume 4f electrons to be well localized in PrFe$_{4}$P$_{12}$, the difference compared to the La-and Nd- compounds should be ascribed only to the electron scattering by the localized 4f-electrons.  Then, we can hardly expect such a large magnitude of TEP at low temperatures and a difference in TEP at RT compared to LaFe$_{4}$P$_{12}$.  The small sample size could lead to a large error in TEP, however, we confirmed the contribution to be minor in the present case from independent measurements on different samples.
	
Figure 8(b) shows the T-dependence of TEP for CeFe$_{4}$P$_{12}$, where the huge value of TEP should be noted.   At around 65K, TEP exhibits a broad peak of $\sim$0.76mV/K and sharply decreases below 50K where $\rho(T)$ shows an apparent increase after a plateau.  The temperature dependence is weak above 100K, except a small but an apparent anomaly at $\sim$220K.  The positive sign of TEP as well as $R_{H}$ at high temperatures is consistent with the band structure calculation.$^{21}$  The top valence bands are dominated by the hybridized Fe-3d and P-3p states with a moderate dispersion, while the lowest conduction bands are dominated by the narrow spin-orbit split Ce-4f bands. Namely, the effective mass for the electrons in the 4f-dominated bands is apparently heavier than that for the holes in the valence bands, which leads to the positive TEP and $R_{H}$ in the intrinsic regime.  

\section{Conclusions} 
All the transport properties of CeFe$_{4}$P$_{12}$ show unusual temperature dependence below RT; the complex temperature dependence of resistivity unexpected for a simple single gap semiconductor, the huge thermoelectric power above $\sim$10K with a maximum value of 0.76mV/K, the sharp decrease of both TEP and the carrier mobility below a peak around 65K.  All these features might be related with the 4f-electron hybridization associated with the filled skutterudite structure as was already pointed out based on the electrical resistivity and optical measurements.

The most remarkable finding was made on PrFe$_{4}$P$_{12}$.  All the transport properties $\rho(T)$, $R_{H}(T)$ and $S(T)$ indicates a strong possibility of Kondo-like scattering.  Ordinary magnetic Kondo effect may not be its origin.  In order to clarify the origin, further investigations such as neutron scattering and ultrasonic experiments are in progress.

The negative d$\rho$/d$T$ above $T_{C}$ in NdFe$_{4}$P$_{12}$ has a magnetic origin, which might be related with an unusual dispersion of the magnon spectrum inferred from the temperature dependence of the resistivity and the specific heat below $T_{C}$.    

\section{Acknowledgments}
The authors are grateful to thank Profs. Y. Onuki, H. Harima, O. Sakai for the helpful discussions.  This work was partly supported by a Grant-in-Aid for Scientific Research from the Ministry of Education, Science, Sports and Culture of Japan.


\section*{Figure Captions}
\begin{figure}[htbp]
\caption{Temperature dependence of electrical resistivity for PrFe$_{4}$P$_{12}$.  The inset shows the $\rho(T)$s on four different samples normalized at the peak, where the difference is only barely discernible.}
\caption{Comparison of the experimental $\rho(T)$ for PrFe$_{4}$P$_{12}$ with the temperature dependence expected for the scattering with a gap structure.}
\caption{Temperature dependence of resistivity for NdFe$_{4}$P$_{12}$.  The inset shows $\rho$ versus $T^{4}$ plot.}
\caption{Temperature dependence of resistivity for four different CeFe$_{4}$P$_{12}$ samples.  The inset shows $\log(\sigma)$ versus $T^{-1}$ plot for \#4.}
\caption{Temperature dependence of the Hall coefficient (a) for La-, Pr-, and NdFe$_{4}$P$_{12}$ and (b) for CeFe$_{4}$P$_{12}$.}
\caption{Comparison of experimental $R_{H}$ for with the skew scattering contribution $\sim M\rho$. (a) for PrFe$_{4}$P$_{12}$ and (b) for NdFe$_{4}$P$_{12}$. }
\caption{Temperature dependence of the Hall mobility in CeFe$_{4}$P$_{12}$ calculated by using $\rho(T)$ and $R_{H}(T)$ data.}
\caption{Temperature dependence of thermoelectric power (a) for La-, Pr-, and NdFe$_{4}$P$_{12}$ and (b) for CeFe$_{4}$P$_{12}$.}
\end{figure}

\end{document}